\documentclass[a4paper,12pt]{article}
\usepackage[utf8]{inputenc}
\usepackage[T2A]{fontenc}
\usepackage{indentfirst}
\usepackage{amsmath}
\usepackage{amsthm}
\usepackage{amsfonts}
\usepackage{amssymb}
\usepackage{graphicx}
\usepackage{setspace}
\usepackage{array}
\usepackage{float}
\usepackage{url}
\usepackage{multirow}
\usepackage{multicol}
\usepackage{enumitem}
\usepackage[title]{appendix}
\usepackage{titling}
\usepackage{longtable}
\usepackage{amsbib}
\usepackage{geometry}
\newtheorem{theorem}{Theorem}

\newtheorem{definition}{Definition}

\providecommand{\keywords}[1]{
	\begin{flushleft}
	\small\textbf{Keywords: } #1
	\end{flushleft}
	}
\providecommand{\affili}[1]{
	\begin{center} 
	\small #1
	\end{center}
	\vspace{.1cm}}
\title{Entropically secure cipher  for messages generated by Markov chains with unknown statistics} 

\author{\normalfont Boris Ryabko$^{1,2}$ 
}

\date{}
\begin{document}
\large
%
%
\maketitle
%
%
\par\vspace{-60pt}
%
\affili{
	$^1$Federal Research Center for Information and Computational Technologies
, Novosibirsk, Russia\\
	$^2$Novosibirsk State University, Russia\\
	\texttt{boris@ryabko.net}
}

%
%
\begin{abstract}
In 2002, Russell and Wang proposed a definition of entropic security, which was developed within the framework of secret-key cryptography.  An entropically secure system is unconditionally secure,  that is, unbreakable regardless of the adversary’s computing power.  In 2004, Dodis and Smith further developed 
 the results of Russell and Wang and, in particular, stated that the notion of an entropically secure symmetric encryption scheme is extremely important for cryptography because one can construct entropically secure symmetric encryption schemes with keys much shorter than the length of the input, thus circumventing Shannon’s famous lower bound on key length. 

   In this report we suggest an entropically secure scheme for the case where  the  encrypted message is generated by a Markov chain with unknown statistics. The length of the required secret key is proportional to the logarithm of the message length (as opposed to the length of the message itself for the one-time pad). 

\end{abstract}
%
%
%
%
\section{Introduction}

In 1949, K. Shannon, in his remarkable article \cite{sh}, described the perfect secret system and showed that the one-time pad is such  a system. Since then, it has been generally accepted that the length of the secret key should be equal to the length of the encrypted message (or at least its entropy).  Russell and Wang  \cite{fool} proposed the notion of 
 entropic security, which gives a possibility to build a symmetric encryption scheme with  a secret key much shorter than the length of the input,   thus, in a sense,  circumventing the mentioned Shannon’s   lower bound on key length.  Informally, the entropy-secure symmetric encryption scheme uses the entropy of the input message to make the required secret key shorter.

 The concept of entropic security has been generalized and developed by Dodis and Smith \cite{do} and  investigated by several other authors \cite{fo2,jr,rya}. 
In order to describe it, suppose that there is a sender Alice and a receiver Bob who share a secret
key $K$,  and Alice wants to securely send some message $M$ to Bob over a public channel. The message $M$ is assumed
to come from some a-priori distribution on ${\Lambda}^n$  where $\Lambda$ is a finite alphabet,   $n \ge 1$, and  $K$ is a sequence of equally probable and independent binary digits. Informally, the goal is to compute  $E(M,K)$ which allows Bob to extract $M$ from $E(M,K)$ using $K$ and (the decoder)  $D(E,K),  (D(E,K) =M),$ in such a way as to  
  reveal ``no information'' about $M $ to the adversary Eve beyond what she already knew.  It is assumed that $E(M,K)$ is a probabilistic map, that is, it can also use random numbers, which are unknown to Bob.

 The following formal definition of the entropic security belongs to Russell and Wang  \cite{fool} (see also 
  Dodis and Smith \cite{do}): 
  \begin{definition}{ }
A probabilistic map $E(M,K)$ is said to hide
all functions $f$ on $\Lambda^n$ to $\{0,1\}^*$
 with leakage $\epsilon, \epsilon > 0,$ if, for every adversary $A$, there exists some adversary $\hat{A}$ (who does not know $ E(M,K)$) 
such that for all functions $f$ from  $\Lambda^n$ to $\{0,1\}^*$, 
\begin{equation}\label{entrsec} 
| \, Pr\{A( E(M,K) ) \, = f(M) \}  \, - Pr\{  \hat{A}(\,)\, = f(M) \} \, | \,\le\, \epsilon .
 \end{equation}
 (Note that $\hat{A}$ does not know $ E(M,K)$  and, in fact, she guesses the meaning of the function $f(M)$, ignoring $E(M,K)$.) 
 
 The cipher $ E(M,K)
 $ is $\epsilon$-entropically secure for a probability distribution $P$ on $\Lambda^n$  if $E(M,K)$  hides all functions  $f$ on $\Lambda^n$ to $\{0,1\}^*$ with leakage $\epsilon$ when $M$ obeys the distribution $P$.  
\end{definition}{ }

Another concept, namely, that of  indistinguishability, provides another way evaluate 
 the strength of the cipher.
To describe it, we first need to define min-entropy.

For a probability distribution $P$ on the alphabet $S$  
the min-entropy  is defined as follows:
 \begin{equation}\label{entr}
h_{min}(P) =  - \log \, \, \max_{a \in S }\, P(a) \, ,
 \end{equation}
 $\log = \log_2$.
 
 \begin{definition}\label{indis} (Dodis and Smith \cite{do}.) 
 A randomized map $Y ()$ is $(t, \epsilon)$-indistinguishable if there is a random variable $ G$
such that for every distribution on a set $\bf{M}$   with min-entropy at least $t$, we have
$$SD(Y (M), G)  \le \epsilon,$$
where for two probability distributions $A,B$  
$$ SD(A,B) = \frac{1}{2} \sum_{M \in \bf{M}}  | Pr\{A = M\} - Pr\{B = M\} | \, .
$$
  \end{definition}
Informally, in what follows the map $Y(\,)$ will be the cipher and, again,
 $G$ does not depend on  the ciphered message. So,  Eve can guess the message regardless of its cipher.

Dodis and Smith \cite{do} showed that entropy security and indistinguishability are equal (up to small constants in key length). 
In particular, they show that if a cipher is $\epsilon$-entropically secure, it is $4\epsilon$-indistinguishable. 

The main result of this paper is as follows: 
We describe an $\epsilon$-entropically  secure cipher for the case where the probability distribution $\mu$ is unknown, but it is known that it belongs to class of stationary ergodic Markov chains with finite memory, or connectivity, $m$, $m \ge 0$, whose definition is given in Appendix.
 (If $m=0$ then the  symbols generated by $\mu$ are  independent and identically distributed~--  i.i.d.). The length of the required secret key is $c_1 \log n + c_2 \log (1/\epsilon) +c_ 3$, where $n$ is the length of encrypted sequence, $c_ 1, c_2$ and $c_3$ are constants that depend on $m$ and the size of the alphabet $\Lambda$. 
 (Recall that all participants know $m$, but  the secret key are known only to Alice and Bob and the key is used only once).

The proposed method is based on the concept of the $\epsilon$-entropically secure cipher and some results of  universal coding, which makes it possible to efficiently ``compress'' messages with   unknown  statistics \cite{kr}. 
\section{Preliminaries}

\subsection{Universal coding}

First, we consider the simplest case where the alphabet is $  \{0,1\}^n, n\ge 1$ and letters are generated by some i.i.d.\ source $\mu$ and $\mu(0)$, $\mu(1)$ are unknown. The goal is to build a  lossless code which ``compresses''  $n$-letter sequences in such a way that the average length (per letter)  of the compressed sequence is close to the Shannon entropy  $h(\mu)$, which 
 is the lower limit of the codeword length   (lossless code is such  that the encoded messages can be decoded without errors and  $h(\mu) = - (\mu(0)\log \mu(0) +
(1-\mu(0)) \log (1-\mu(0) )   $ ) \cite{kr,co}.  

The first universal code was  invented by Fitingoff \cite{fi} and we use this code as a part of the suggested entropically secure cipher.  In order to describe this code we consider any word $v\in \{0,1\}^n$ and 
denote by $\nu$ the number of ones in  $v  $ and let $S_\nu$ be the set of n-length words with $\nu$ ones.
Fitingoff  proposed to encode the word $v $ by two subwords $u$ (prefix) and $w$ (suffix), where $u$ is the binary notation of an integer $\nu$   and $w$ is the index of the word $v$ in the subset $S_\nu$. It is assumed that the words in $S_\nu$ are ordered 0 to $(|S_\nu| -1)$ (say, lexicographically) and the lengths of $u$ and $w$ are equal to $\lceil \log (n+1) \rceil $ and $\lceil \log |S_n| \rceil $, respectively.
For example, for $n= 3$, $v = 100$  we obtain $\nu = 1, u = 01, w= 10$.

Recall the definition of the so-called  prefix-free code.
A  set of words $U$ is  prefix-free if for any  $u,v \in U$ neither $u$ is a prefix of $v$ nor $v$ is a prefix of $u$ \cite{co}.  Clearly, the Fitingoff code is prefix-free. 
If some code $ \lambda$ is prefix-free, then  for any sequence $x_1 x_2  .... x_n, n\ge 1,$   $x_i \in \Lambda$,  the encoded sequence   $\lambda(x_1) \lambda(x_2) ...  \lambda(x_n) $ can be decoded to $x_1 x_2  .... x_n$ without errors.   Hence, any prefix-free code is a lossless one. 

 If we denote the Fitingoff code  by $code_F$ we obtain from its description
\begin{equation}\label{fi1} 
|code_F(v) | =  \lceil \log (n+1) \rceil  + \lceil \log |S_\nu| \rceil  \, +1 \,.
\end{equation}
For this code the ability to compress messages is based on the simple observation that probabilities of all messages from $S_\nu$ are equal  for any distribution $\mu$ and, hence, $\,\, \mu(v )\le 1/|S_\nu|$ for   $\mu$ and any word $v \in S_\nu$. From this  inequality and (\ref{fi1}) we obtain 
\begin{equation}\label{fi2}
|code_F(v) | \le    \log (n+1)   + 3 +
\log(1/\mu(v) ) \, .
\end{equation}
(Let's explain the name ``universal code.''  Clearly, the average code-length $E_\mu(|code_F|)$  is not grater than $ \log (n+1)   + 3  + n h(\mu)$ and, hence,
the average length per letter $E_\mu(|code_F|)/n$  is not grater than $h(\mu) + (\log n +3)/n$).  We can see that     $E_\mu(|code_F|)/n$ $\to$ $h(\mu)$ if $n \to \infty$. So, one code compresses sequences generated by any  $\mu$, that is, the code universal.)

The Fitingoff code described generalizes to i.i.d. processes with any finite alphabet $\Lambda$, as well as to Markov chains with   memory or connectivity $m$,  based on the same method as for binary i.i.d. \cite{kr}.  Namely, the set of all $n$-letter words is divided into subsets of equiprobable words, and the code of any word is represented by a prefix and a suffix, 
where the prefix contains the number of the set with equiprobable words which contains the encoded one, 
and the prefix is  the  number in this set.  It can be shown that the number of sets with 
equiprobable words is bounded above by 
$(|\Lambda |-1) |\Lambda |^{m}$ (\cite{kr,co}),
 and similarly (\ref{fi2}) we can deduce that
\begin{equation}\label{fi3}
|code_F(v) | \le \log ( (|\Lambda|-1) |\Lambda|^{m} ) + 3 +
 \log(1/\mu(v) ) \, .
\end{equation}

It is important to note that there exists an algorithm to find  the codewords which is based on method of fast calculation of numbers in $S_\nu$, see \cite{rya2}. The complexity of this algorithm is $O(n \log^3 n \log \log n)$.  

\subsection{Entropically secure ciphers}


 Dodis and Smith \cite{do}, based on the results of Russell and Wang  \cite {fool}, proved the following 
 
 {\bf Theorem (Russell-Wang,  Dodis- Smith ) } (\cite {fool}, \cite{do}).
 Let there be a probability distribution $\sigma$ on an alphabet $\Lambda = \{0,1\}^l$, $l \ge 1$.
Then, for any  $\epsilon >0$, there exists an $\epsilon$- entropically  secure cipher $E(M,K)$, $M \in \{0,1\}^l$ 
with the length of the key 
\begin{equation}\label{ds}
|K|=  l -  h_{min}(\sigma)+ 2  log (1/\epsilon) +2.
\end{equation}

Take any such  cipher and denote it $cipher_{RW-DS}(M,K)$. 
Dodis and Smith described three algorithm of such  ciphers   with a key length (\ref{ds}) whose complexity grows polynomially in $l$ and $\log (1/\epsilon)$ (One such a cipher is described in Appendix).

It is important to note that each 
of the three constructions of the ciphers depends only on min-entropy, that is, the cipher construction is the same for all distributions with the same 
 min-entropy (but, of course, depends on $\epsilon$ and $l$).

\section{The cipher}
\subsection{Randomised prefix-free codes}

Let $\lambda$ be a prefix-free code for some alphabet  $\Lambda^*$ and $L = \max_{a \in \Lambda^*}|\lambda(a)| \, .$
The randomized code $\rho_\lambda$  
 maps elements from $\Lambda^*$  to the set 
$\{0,1\}^L$ defined as follows:
\begin {equation}\label{ro-la}
\rho_\lambda(a_i) = \lambda(a_i) \,
    r^i_{|\lambda(a_i)|+1} r^i_{|\lambda(a_i)| + 2} ... r^i_L \, , \end{equation} 
where  $r^i_{|\lambda(a_i)|+1}, r^i_{|\lambda(a_i)| + 2}, ... ,r^i_L$  are uniformly distributed and independent random bits (for all $i$).

Let us define the probability distribution $\pi_{\lambda,\mu}$ on   $\{0,1\}^L$ as follows: 
\begin {equation}\label{rand}
\pi_{\lambda,\mu}(y_1y_2 ... y_L) =
   \mu(a) 2^{- (L - |\lambda (a) | )} \,   $$ $$     \text{if}  \quad
    y_1 y_2 ... y_{|\lambda(a_i)|} = \lambda(a). 
    \end{equation} 
    If  for some  $y=y_1 ... y_L$  any $\lambda(a)  $ is not a prefix of $y$, then $\pi_{\lambda,\mu}(y) = 0$. 
    
    Let us estimate the min-entropy of the distribution $\pi_{\lambda,\mu}$. From this equation and the definition of the min-entropy (\ref{entr}) we obtain  the following:
    
\begin {equation}\label{claim}
h_{min}(\pi_{\lambda,\mu}) = L - \max_{a \in \Lambda} (|\lambda(a) | - \log (1/\mu(a) ) \, .
\end{equation}

Now we consider the Fitingoff code applied to $n$-letter sequences generated by a Markov chain $\mu$  of memory $m$ over 
 some alphabet $\Lambda$. 
The Fitigoff code is prefix-free and, hence,        from (\ref{fi3}) and (\ref{claim}) we obtain  
the following 

{\bf Statement.} For any distribution $ \mu$

\begin {equation}\label{minSh}
h_{min}(\pi_{code_F, \mu}) >   L  - (|\Lambda|^m (|\Lambda| - 1) \log n  \, + \,3)\, .
\end {equation}
       In particular, for an i.i.d. source with binary alphabet 
   $$
h_{min}(\pi_{code_F}) >   L  - (\log n +3)\, .
$$

\subsection{Description of the cipher}    
Here we describe a cipher with the key of length   $const_1  \log n + const_2\, \log (1/\epsilon ) + const_3 $, which is   $\epsilon$-entropically secure for $n$-letter sequences generated by any 
 (unknown)  Markov chain $\mu$  of memory $m$ over some alphabet $\Lambda$. 

Briefly, the encryption is done as follows: first compress the message with the Fitingoff code, then randomize the encoded message according to (\ref{ro-la})  and then encrypt the received $ \rho_{code_F, \mu}(\,)$ with an entropically secure cipher. 
(Note that the distribution of $\mu$  is unknown.) 

In detail, this algorithm is as follows:

 {\bf Parameters:}  $\epsilon > 0$,  the alphabet $\Lambda $,  the memory of  Markov chain $m$  and the length of the ciphered message $n$.

{\bf Input:}  a word $v \in \Lambda^n$.

{\bf 1st step:}  Encode $v$ with the Fitigoff code  $code_F(v)$ (with parameters $\Lambda,m$ and $n$).

{\bf 2nd step:}  Calculate the random word  $\rho_{code_F}(v)$ ($ \in \{0,1\}^L$).

{\bf 3rd step:} Calculate the $\epsilon$-entropically  secure  cipher 
 $cipher_{RW-DS}(\rho_{code_F}(v),K) $  with the length of the secret key $|K|=  (|\Lambda|^m (|\Lambda| - 1)  \log n + 2 \log (1/\epsilon) + 5$ bits.

{\bf Output:} 
$cipher_{RW-DS}(\rho_{code_F}(v))$. 

The decryption algorithm is as follows:
first Bob decrypts  the word $E(\rho_{code_F}(v),K )$  ($=cipher_{RW-DS}(\rho_{code_F}(v)) \,$)  with the known secret key $K$ and obtains the word $\rho_{code_F}(v)$. Then, based on the prefix-free property of the Fitingoff code, Bob finds 
 the word $code_F(v)$ and then decodes it to get $v$.

 The described cipher uses compression and randomisation.  Denote it  $cipher_{c\&r}$.

   The theorem  of Russell-Wang  and   Dodis-Smith  guarantees the entropic security and indistinguishability for the first cipher $cipher_{RW-DS}$, so, we need to prove a similar property for  the proposed $cipher_{c\&r}$.  Despite the equivalence of the concepts of entropic security and indistinguishability \cite{do}, we will prove these properties separately due to the great importance of this fact for the described cipher $cipher_{c\&r}$.
   
                The following theorem describes the entropic security property for this cipher:
\begin {theorem}\label {my}  
Let $\epsilon >0$ and suppose that the cipher $cipher_{c\&r}$ is applied to  
 $n$-letter words $M$ generated by  a stationary ergodic Markov chain with memory $m, m\ge 0$, and  an alphabet $\Lambda$,  and let the length of the secret key $K$ be $  (|\Lambda|^m (|\Lambda| - 1)  \log n + 2 \log (1/\epsilon) + 5$.
Then 
  $cipher_{c\&r}$ is $ \epsilon$-entropically secure,  
that  is, 
for any function $A : \, \{0,1\}^L \to \{0,1\}^*$ and $f: \Lambda^n \to \{0,1\}^*$ 
there exists such a function $\hat{A} : \, \{0,1\}^L \to \{0,1\}^*$ that
$$
| Pr \{ A(cipher_{c\&r} (M,K )  = f(M) \} - Pr \{\hat{A}(\,) = f(M) \} | \le \epsilon ,
$$
where $\hat{A}$  does not use $cipher_{c\&r} (M) $.
\end{theorem}
{\it Proof.}   
The cipher $cipher_{RW-DS}(\rho_{code_F}(v),K) $  with the length of the secret key $|K|=  (|\Lambda|^m (|\Lambda| - 1)  \log n + 2 \log (1/\epsilon) + 5$ is applied to $\{0,1\}^L$  (see the step 3).
First we note  that the cipher is $\epsilon$-entropically secure. Indeed, from Theorem  of Russell-Wang and  Dodis- Smith  (see (\ref{ds}))   and the estimate of the min-entropy (\ref{minSh})   we can see that such a cipher exists for the distribution $\pi_{code_F,\mu} $  for any (unknown) $\mu$.  So, from the definition of $\epsilon$-entropical  security we can see that  for any function $g$
$$ 
 | Pr \{ A(cipher_{RW-DS}(v)  = g(v) \} - Pr \{\hat{A}(\,) = g(v) \} | \le \epsilon,
$$
where $v, v \in \{0,1\}^L$, 
$g$ is any function defined on $\{0,1\}^L$ ($g:  \{0,1\}^L\to \{0,1\}^*)$ and $\hat{A}(\,)$ does not depend on $v$ (to be short, $\lambda = code_F$).
Taking into account that the code $\lambda$ is prefix-free,  we can define such a function $\phi$  that 
for any  $a \in \Lambda^n$ and  $u=\rho_\lambda(a) $,    $\,\, \phi(u) = a$.
For any function $f:  \Lambda^n\to \{0,1\}^*$  and $M$ consider the function 
$g(\rho_\lambda(M)) = f(\phi (\rho_\lambda(M)) (=f(M))$. This equation is valid for the  function $g$ and  for $v= \rho_\lambda(M)$, hence 
$$ 
 | Pr \{ A(cipher_{ds}(\rho_\lambda(M) )  =  f(\phi (\rho_\lambda(M)) \} - $$ $$ Pr \{\hat{A}(\,) =  f(\phi (\rho_\lambda(M))  \} | \le \epsilon .
$$
  Taking into account that  $cipher_{c\&r} (M) = cipher_{RW-DS}(\rho_\lambda(M) )$
   and $ f(\phi (\rho_\lambda(M))  = f(M)$, we can see from the latter inequality that  
   $$ 
 | Pr \{ A(cipher_{c\&r}( M))  = f(M) \} - $$ $$ Pr \{\hat{A}(\,) = f(M) \} | \le \epsilon \, .
$$
The theorem is proven.

The following theorem establishes   indistinguishability  of 
 $cipher_{c\&r}$.
\begin {theorem}\label {my2}  
Let $\epsilon >0$ and  and suppose that the cipher  $cipher_{c\&r}$ is applied to $n$-letter words $M$ generated by  a stationary ergodic Markov chain with memory $m, m\ge 0$, and  an alphabet $\Lambda$,  and let the length of the secret key $K$ be $  (|\Lambda|^m (|\Lambda| - 1)  \log n + 2 \log (1/\epsilon) + 5$. Then, this cipher is $4\epsilon$- indistinguishable.
 \end {theorem}
{\it Proof.} 
The cipher $cipher_{RW-DS}$ is $\epsilon$-entropically secure (see Theorem 1). As we mentioned in Introduction, Dodis and Smith  \cite{do}  showed that it means that this cipher is $4 \epsilon $-indistinguishable. Our goal is to prove this property 
 for $cipher_{c\&r}$.
The $4 \epsilon $-indistinguishability means that
$SD(cipher_{RW-DS}  , G) \le 4 \epsilon $, where $G$ is a random variable on $\{0,1\}^L$ (which  is independent on $cipher_{RW-DS}  $).
  
Define $U_a = \{ cipher_{RW-DS}  (\lambda(a)  \, r ): r \in \{0,1\}^{L-\lambda(a)} \}$ 
and let  the  a random variable  of $G'(v) $ be defined as follows:
$$Pr\{G' = v\}  = \sum_{w \in U_v} Pr\{ G = w\}.$$
The following chain of equalities and inequalities is based on these definitions and the  triangle inequality for $L_1$:
$$SD(cipher_{c\&r}, G') =  $$ $$\frac{1}{2 }\sum_{u \in \Lambda^n } | Pr\{cipher_{c\&r}=u\} - Pr\{ G'=u \}| =
$$
$$  \frac{1}{2 }\sum_{v \in \{0,1\}^n }| \sum_{w \in  U_v } (
 Pr\{cipher_{RW-DS} =w\} - Pr\{ G=w \} )| \le
$$
$$ \frac{1}{2 }\sum_{v \in  \Lambda^n } \sum_{w \in  U_v }|
Pr\{cipher_{RW-DS} =w\} - Pr\{ G=w \} \, | =
$$
$$
\frac{1}{2 } \sum_{w \in \{0,1\}^L} 
Pr\{cipher_{RW-DS} =w\} - Pr\{ G=w \} \, |  = 
$$
$$
SD(cipher_{RW-DS} , G) \le  4\epsilon \, .
$$
So, $SD(cipher_{c\&r}, G') \le 4  \epsilon $.

Theorem is proven.

Let us estimate the complexity of encoding and decoding. As we mentioned above, the encoding and decoding fitting complexity is $O(n \log^{const} )$. The complexity of the Dodis and Smith cipher is polynomial in $n$. Thus, the complexity of the proposed cipher is also polynomial in $n$.

\section{Appendix} 
\subsection{  The definition of a stationary ergodic Markov chain with memory, or connection, $m$.}       First we give a   definition of stationary ergodic processes.
  The time shift $T$ on $\Lambda^\infty$ is
defined as $T(x_1,x_2,x_3,\dots)=(x_2,x_3,\dots)$. A process $P$
is called stationary if it is $T$-invariant: $P(T^{-1}B)=P(B)$ for
every Borel set  $B\subset \Lambda^\infty$. A stationary process is
called ergodic if every $T$-invariant set has probability 0 or 1:
$P(B)=0$ or $1$ whenever $T^{-1}B=B$  \cite{bi,D}.

We denote by $M_\infty(\Lambda)$  the set of all stationary and ergodic
sources  and let $M_0(\Lambda) \subset M_\infty(\Lambda)$ be the set of all
i.i.d. processes. We denote by  $M_m(\Lambda) \subset M_\infty(\Lambda)$ the
set of Markov sources of order (or with memory, or connectivity)
not larger than $m, \, m \geq 0.$  By definition $\mu \in M_m(\Lambda)$
if
$$
\mu (x_{t+1} = a_{i_1} | x_{t} = a_{i_2}, x_{t-1} = a_{i_3},\,
...\,, x_{t-m+1} = a_{i_{m+1}},... ) 
$$
 $$ = \mu
(x_{t+1} = a_{i_1} | x_{t}  = a_{i_2}, x_{t-1} = a_{i_3},\, ...\,,
x_{t-m+1} = a_{i_{m+1}}) $$ for all $t \geq m $ and $a_{i_1},
a_{i_2}, \ldots \, \in \Lambda.$ 

\subsection{Entropically secure ciphers.}
In this part we describe one entropically secure cipher from \cite{do}, part 3.2.

Let $\{ h_i \}_{i \in I}$ be some family of functions $ h_i : \{0, 1\}^k \to  \{0, 1\}^n$, indexed over the set $I = \{0, 1 \}^r$. 
By definition, a collection of functions from $n$-bit words to $n$-bits   is XOR-universal if: $$ \forall a, x, y \in \{0,1\}^n, x \neq y,  Pr \{h_i(x) \oplus  h_i(y) = a \} \le \frac{1}{2^{n-1}} \, , 
$$
if $i$ is randomly chosen from $I$ according to the uniform distribution ($\oplus$ is symbol-by-symbol modulo 2 summation). 
Also, suppose that  there is a   XOR-universal collection of functions  whose 
 description    is public and, hence, it
  is known to Alice, Bob and Eve. 
  
Dodis and Smith  consider an encryption scheme of the form
$$ E(m,K,i) = (i; m  \oplus h_i(K)$$
where   $i$ is randomly chosen from $I$ according to the uniform distribution, and $K$ is a  $k$-bit secrete key. Note that $m$ is a ciphered message  of  length $n$, $i$ is the number of $h_i$ in the set $I$ and $i$ $=\log  | I | = r$. 
(Dodis and Smith notice that this scheme is a special low-entropy, probabilistic one-time pad.)  Decryption is 
obviously possible, since the description of the function $h_i$  is public.  It is shown \cite{do} that this cipher is $\epsilon$-entropically secure for 
$  |k| \ge
n - h_{min} + 
2 \log (1/\epsilon) +2
$ 
 if the function family $\{h_i\}_{i\in I }$
is XOR-universal.

An example of XOR-universal family is as follows   \cite{do}:
View $\{0, 1\}^n $ as $\mathcal{F} = GF(2^n)$, and embed the key set $\{0, 1\}^k$ as a subset
of $\mathcal{F}$.
For any  $i   \in  \mathcal{F}$, let $h_i(K) = i K$, with multiplication in $\mathcal{F}$. This yields a family of linear maps
$\{h_i\} $ with $2^n$ members.
For this family the complexity of ciphering and deciphering  is $O(n \log n \log\log n)$ \cite{do}. 

It is important to note that the length of the secret key ($k$) depends only on the min-entropy of the probability distribution and does not depend on other  parameters of the
  distribution.

%
%
\def\refname{References}

%
%
\end{document}